# Planar chiral metasurfaces with maximal tunable chiroptical response driven by bound states in the continuum

Tan Shi[1,†], Zi-Lan Deng[1,†,*], Guangzhou Geng[2,†], Yixuan Zeng[3], Guangwei Hu[3], Adam Overvig[4], Junjie Li[2,*], Cheng-Wei Qiu[3], Andrea Alù[4], Yuri S. Kivshar[5], and Xiangping Li[1,*]

[1]Guangdong Provincial Key Laboratory of Optical Fiber Sensing and Communications, Institute of Photonics Technology, Jinan University, Guangzhou 510632, China

[2]Beijing National Laboratory for Condensed Matter Physics, Institute of Physics, Chinese Academy of Sciences, Beijing 100191, China

[3]Department of Electrical and Computer Engineering, National University of Singapore, Kent Ridge 117583, Republic of Singapore

[4]Photonics Initiative, Advanced Science Research Center, City University of New York, New York, NY 10031, USA

[5]Nonlinear Physics Center, Australian National University, Canberra ACT 2601, Australia

[†]These authors contributed equally to this work
[*]E-mail: zilandeng@jnu.edu.cn; jjli@iphy.ac.cn; xiangpingli@jnu.edu.cn

**Abstract:**
Optical metasurfaces with high-Q chiral resonances can boost light-matter interaction for various applications of chiral response for ultrathin, active, and nonlinear metadevices. Usually, such metasurfaces require sophisticated depth-resolved nanofabrication to realize subwavelength stereo-nanostructures, posing overwhelming challenges, especially in the short-wavelength range. Here, we suggest a novel planar design for chiral metasurfaces supporting bound states in the continuum (BICs) and demonstrate experimentally chiroptical responses with record-high Q-factors (Q=390) and near-perfect circular dichroism (CD=0.93) at optical frequencies. The symmetry-reduced meta-atoms are highly birefringent and support winding elliptical eigen-polarizations with opposite helicity surrounding the BIC polarization singularity, providing a convenient way for achieving maximal planar chirality tuned by either breaking in-plane symmetry or changing illumination direction. Such sharply resonant chirality realized in planar metasurfaces promises various practical applications in classical and quantum optics including chiral sensing, enantiomer selection, and chiral quantum emitters.

**One Sentence Summary:** This work employs the physics of chiral BIC with strong birefringence and experimentally demonstrates the first of planar chiral metasurfaces at optical frequencies with simultaneously mediated near-perfect circular dichroism (CD=0.93) and ultrahigh Q-factors (Q=390).

**Introduction**

Chirality refers to specific property when an object cannot be superimposed with its mirror after rotation or translation[1-3], which generically exists in nature in organic molecules[4], quartz crystals[5], and many others. The study of chirality is fundamentally important in various areas including analytical chemistry, pharmaceutics, and even searching for extraterrestrial life. The light interaction with those geometries can render chiroptical effects including circular dichroism (CD) and optical activity, manifested by a difference in intensity and phase responses between left/right circularly polarized (LCP/RCP) light illuminations. Optical chirality was first demonstrated in three-dimensional (3D) photonic structures[6] such as the helices[6,7], twisted cross structures[8,9] and multi-layered structures[10-12], all exhibiting strong CD and optical activity with preserved circular polarization, which however require very demanding 3D nanofabrication techniques. Recently, two-dimensional (2D) or planar structures have been shown to support both intrinsic planar chirality at normal incidence[13-16] and extrinsic chirality at oblique incidences[17-21]. Being different from 3D chirality with broken mirror symmetry in the propagation direction, although even existing in some stacked structures[15], planar chirality shows a circular polarization conversion between the output and incident light, and the cross-polarization behavior can be further exploited to modulate geometry phase for arbitrary chiral wavefront shaping[22-26]. However, it is still challenging to achieve maximal planar chirality simultaneously with ultra-high quality (Q-) factor due to absorption and scattering loss in previous approaches, hindering many active applications that rely on extremely strong chiral light-matter interaction.

Importantly, bound states in the continuum (BICs) can provide a feasible solution for the outstanding problems of chiral photonics. BICs are identified as localized states coexisting with extended modes within light cones, and they have attracted tremendous attention due to their infinite Q-factors, thus significantly boosting light-matter interaction for many applications such as surface-emitting lasing[27,28], biomedical sensors[29-31], and nonlinear frequency conventers[32-36]. BICs originate from destructive interference of several radiative channels, and they can be formed by either symmetry protection[37-40] or accidental resonances[41-47]. Since perfect BICs are nonradiative and cannot be accessed externally, the quasi-BICs (q-BICs) with high-Q resonances are tailored via perturbing the symmetry for practical realizations. Recently, chiroptical nanostructures mediated by BIC have been proposed, rendering perfect unitary chirality and extremely high Q-factors[48,49], and greatly expanding technology for achieving optical chirality[49-54]. However, such chiral q-BIC scheme has only been proposed by complicated stereo-nanostructures so far, making it elusive for experimental realization and application development especially at optical frequencies.

Here, we suggest and realize experimentally a novel design of planar chiral metasurfaces at optical frequencies with simultaneously mediated maximal chirality and ultrahigh Q-factors. A double-sided scythe (DSS-) shaped α-Si inclusion with in-plane inversion $C_2$ symmetry but without in-plane mirror symmetry is employed to construct the BIC state. Such BIC state with strong birefringence is a vortex polarization singularity (*V* point) surrounded by elliptical eigenstate polarizations with nonvanishing helicity, providing a convenient way to achieve maximum and high-Q chirality by slightly perturbing the inversion symmetry. With either breaking in-plane inversion symmetry or varying illumination geometry, we realize planar chiral q-BIC states with extremely strong intrinsic or extrinsic chirality exceeding CD=0.99 (in theory) and CD=0.93 (in experiment). For the intrinsic chirality, the Q-factor of the CD spectra has an inverse quadratic relation with respect to the geometry asymmetry, with a nearly unchanged giant CD peak. For the extrinsic chirality, both the value and sign of CD can be tuned flexibly by changing the incident angles encircling the Γ point, which is connected with the intrinsic nonvanishing helicities of the eigen-polarizations near the BIC singularity[55]. We believe that the demonstrated planar chiral metasurfaces governed by the BIC physics may find many applications in chiral sensing of biomolecules, spin-selective emitters, and active chiroptical devices.

**Results and Discussion**

Previous q-BIC responses in metasurfaces usually were without polarization effect[28,41,42] (Fig. 1a) or with only linear polarization selectivity[29,37,39]. Although very recently q-BICs with versatile polarization responses (chirality selectivity) were theoretically proposed, in form of complicated 3D structures with broken symmetry in the propagation direction[48,49,56] (Fig. 1b), being still a challenge for current 2D patterning technology, especially at optical frequencies. Our present work shows that planar metasurfaces with only reduced in-plane symmetry were able to support q-BIC with both extrinsic and intrinsic planar chirality (Fig. 1c). Figure 1d shows the double-sided scythe (DSS) α-Si structure (refractive indexes refer to Supplementary Fig. S2) employed as the unit cell of our proposed planar metasurface. Such DSS structure ($W_1=W_2$ and $L_1=L_2$, but $L_1 \neq W_1$) has in-plane inversion ($C_2$) symmetry, but lacks any in-plane mirror symmetries. It hosts a BIC state at Γ point characterized by vertical magnetic dipole (MD) mode as shown in Fig. 1e. Different from the Γ point BIC state supported by highly symmetric inclusions[55], which is companied by polarization singularity surrounded by vortex linear polarizations, the BIC state (marked as a black dot in the upper panel of Fig. 1f) of the DSS structure is a polarization singularity enclosed by vortex elliptical eigen-polarizations with non-vanishing helicities (upper panel of Fig. 1f and Supplementary Fig. S3). Such inherent helicity provides a convenient way to achieve strong chirality of scattered light at the vicinity of the Γ point (lower panel of Fig. 1f). Such strong planar chirality behavior can be predicted by temporal coupled mode theory (TCMT) (Supplementary Notes I & II), which suggests that, the perfect planar chiral response can be obtained by tuning the local response to be birefringent using a geometry that has an inversion center, and then breaking inversion symmetry with a suitable perturbation to yield an intrinsic planar chiral q-BIC. While our proposed DSS inclusion with reduced symmetry provides a smart perturbation degree of freedom along the 45° direction, capable of achieving both maximum extrinsic and intrinsic chirality by perturbing a single parameter. As a general remark, in our systems, the sign of the CD is determined by the sign of the helicity, while the maximum of CD does not necessarily coincide with circular eigenpolarization with the highest helicity, because there is non-zero background scattering accumulated by Jones matrix elements (see Supplementary Notes I and Supplementary Fig. S4 ).

Based on such BIC state supported by DSS inclusion with reduced symmetry, one can expect both the extrinsic and intrinsic planar chirality by transferring the BIC to q-BIC. Figures 2(a-d) show the q-BIC with near-unity extrinsic chirality achieved by illumination symmetry breaking (varying incident angle $\theta$ and conical angle $\varphi$). Here, the CD is defined as the transmittance difference under right-handed polarization (RCP) and left-handed polarization (LCP) incidence:

$$CD = \frac{(T_{rr}+T_{lr})-(T_{rl}+T_{ll})}{(T_{rr}+T_{lr})+(T_{rl}+T_{ll})}, \quad (1)$$

where, $T_{ij}=|t_{ij}|^2$ ($i=r, l; j=r, l; r$ represents RCP, $l$ represents LCP) is the transmittance of output polarization $i$ from the input polarization $j$, and all element of $t_{ij}$ construct the Jones matrix under circular polarization basis[11,57],

$$J_{circ} = \begin{pmatrix} t_{rr} & t_{rl} \\ t_{lr} & t_{ll} \end{pmatrix}, \quad (2)$$

Figure 2b shows the calculated transmission spectra of all Jones matrix elements and the CD spectrum of the DSS metasurface with $C_2$ symmetry ($W_1=W_2$ and $L_1=L_2$, but $L_1 \neq W_1$) at oblique incidence ($\theta=8°$, $\varphi=90°$). $T_{lr}$ has a sharp peak greater than 0.9, while all the other three Jones matrix elements ($T_{rl}$, $T_{ll}$ and $T_{rr}$) exhibit dips close to 0 at the resonant wavelength (1456 nm) of the q-BIC, resulting in an ultra-sharp CD spectrum with a maximum greater than 0.95 (green curves in Fig. 2b). Figure 2c shows the evolution of CD spectra by continuous varying incident angle $\theta$ along the $y$-direction ($\varphi=90°$) direction (evolutions of transmission spectra of all Jones matrix elements are shown in Supplementary Fig. S5). The linewidth of the CD spectrum increases from 0 to finite values as $\theta$ increases, and the peak CD sustains near-unitary with finite linewidths, manifesting the transfer process from the non-radiative BIC to high-Q radiative q-BICs with strong extrinsic chirality. The Q-factor of the extrinsic chiral q-BIC shows the standard inverse quadratic law as the illumination asymmetry parameter $\alpha_I=\sin\theta$ (Fig. 2d), which provides a convenient way to tailor the Q-factor of the chiral response. The evolutions of the extrinsic chiral q-BIC along the $x$-direction ($\varphi=0°$) are shown in Supplementary Fig. S6, variation trend is the same as the $\varphi=90°$ case, except that, the sign of CD is flipped due to the opposite helicity of eigen-polarization supported by the DSS structure along $x$- and $y$-directions (Fig. 1f). To investigate the full CD flip picture, the evolution of CD spectra with varying conical angle $\varphi$ at a fixed incident angle $\theta=12°$ is shown in Supplementary Fig. S7. The opposite varying trend of $T_{lr}$ and $T_{rl}$ can be observed as $\varphi$ is varying, leading to the sign flip of CD at positions near $(2n-1)\times45°$, ($n=1,2,3,4$), which is consistent with the prediction provided by Fig. 1f.

Beyond the extrinsic chiral behavior, one can also access intrinsic high-Q chiral behavior under normal incidences. This can be achieved by introducing an in-plane geometry asymmetry parameter

($\delta=W_2-W_1$) as shown in Fig. 2e. As the DSS structure makes strong birefringence along 45°, an in-plane perturbation that breaks the inversion symmetry will provide a necessary degree of freedom to obtain the planar chirality as the TCMT suggests in supplementary Notes II. Figure 2f depicts transmission spectra of all Jones Matrix elements and the CD spectra of the metasurface with $\delta=40$ nm ($W_1=191$ nm, $W_2=231$ nm). The ultra-sharp resonant peak of $T_{lr}$ and resonant dips of $T_{rl}$, $T_{ll}$ and $T_{rr}$ can be obtained at the wavelength of 1392 nm, resulting in a CD peak greater than 0.99 and ultra-narrow linewidth 1.45 nm. Here, we also recognize the planar chiral q-BIC modes by utilizing multipole expansions and nearfield electromagnetic patterns of the modes (Supplementary Fig. S8), which further confirms that the high-Q planar chiral q-BIC states are dominated by the MD mode. Figures 2(g-h) show the CD evolution spectra as a function of the asymmetry parameter $\delta$, the Q-factor of the CD spectrum also manifests an inverse quadratic relationship against the relative asymmetry parameter defined by $\alpha_2=\delta/W_1$, which is the signature of symmetry-protected BIC (corresponding spectra for all components of the Jones matrix spectra refer to Supplementary Fig. S9). Different from the intrinsic 3D chirality with preserved helicity between input and output polarizations, the intrinsic planar chirality is always accompanied by circular conversion dichroism (CCD) and asymmetry transmission (AT)[2,13,58] as shown in Supplementary Fig. S10, due to the preserved symmetry in the propagation direction. The CCD is produced by the mutual orientation of chiral elements and light transmission direction, and the peak value of CCD is close to 1, which is slightly higher than CD (Supplementary Fig. S11). Perfect unitary circular conversion transmittance $T_{lr}$ and CD can be achieved by strictly symmetric background as shown in Supplementary Fig. S12. The existence of a practical $SiO_2$ substrate introduces a perturbation of the background scattering, resulting in reduced transmittance and CD as presented in Figs. 2(b and f).

Due to the symmetry-protected origins of this phenomenon, our designed DSS metasurface has good tolerance against fabrication imperfections as tested in Supplementary Fig. S13, which provides a convenient way for experimental realization. We fabricated the DSS metasurfaces with different in-plane asymmetry parameters ($\delta=0$, 20, 40, 60, and 80 nm), and employed a homemade optical setup (Fig. 3a) to measure the Jones matrix elements of the metasurface at a rotating stage. We first examine the extrinsic chiral q-BIC by measuring the DSS metasurface without in-plane geometry asymmetry ($\delta=0$) for varying incident angles (Fig. 3b). Figure 3c shows the comparison between the simulated and measured transmission Jones matrix spectra at different incident angles ($\theta=0°$, $\pm4°$, $\pm8°$, $\pm12°$) along the $\varphi=90°$ direction. At the normal incidence (middle panel of Fig. 3c), all transmission spectra are smooth without resonant features, because the BIC state at the Γ point is decoupled from any scattering channels of the external environment. At other incident angles, sharp circular conversion peaks of $T_{lr}$ and resonant dips of other three Jones matrix elements ($T_{ll}$, $T_{rl}$, $T_{rr}$) emerge for both simulation and experiment. The measured absolute peak transmittances are greater than 0.78 for all incident angles, which is a little lower than the simulated 0.9, mainly due to the practical scattering loss of the fabricated sample as well as a limited collection efficiency of the measurement setup. As expected, we can clearly observe the linewidth increment of the q-BIC peak with the increasing of incident angle in the experiment, which is consistent with the simulations.

From the measured Jones matrix spectra, we extracted the CD spectra for different incident angles corresponding to Eq. (1) in Fig. 3d. As we can see, sharp CD peaks appear at the q-BIC resonant wavelengths, the simulated/measured CD maxima reach 0.8/0.72, 0.95/0.82, and 0.96/0.88 with the simulated/measured Q-factors of 616/258, 143/96, and 58/39 for incident angles $\theta=\pm4°$, $\pm8°$, $\pm12°$, respectively, while for normal incidence, the CD is always 0 as the BIC state is not accessed by external excitations. We also experimentally characterized the extrinsic chiral q-BIC along the $\varphi=0°$ direction in supplementary Fig. S14 where sharp transmission peaks of $T_{rl}$ are observed, achieving opposite signs of simulated/measured CD: -0.89/-0.82, -0.95/-0.88, and -0.97/-0.91 with the simulated/measured Q-factors of 798/312, 196/102, 85/55 at incident angles $\theta=\pm4°$, $\pm8°$, $\pm12°$, respectively, which further confirms the CD flip behavior.

To further experimentally verify the intrinsic chiral q-BIC, we characterized the DSS metasurfaces with different in-plane geometry asymmetries as shown in Fig. 4. Figure 4a shows the scanning electron microscope (SEM) images of fabricated samples with different asymmetry parameters $\delta$. From both the top view and slant view, we see that the fabricated DSS silicon pillars are smooth and uniform, in satisfactory agreement with the theoretical design. Figures 4(b and c) show the simulated and measured transmission Jones matrix and CD spectra of the five fabricated metasurfaces. When $\delta=0$ (lowest panel in Figs. 4b and 4c), there is no resonance with $T_{rl}=T_{lr}$ and $T_{ll}=T_{rr}$ due to reciprocity, and CD=0 crossing the whole frequency band. As the asymmetry parameter $\delta$ became non-zero, sharp circular conversion transmission peaks of $T_{lr}$ and CD appear, and the bandwidth increases with the increase of the asymmetry parameter (upper four panels in Figs. 4b and 4c). The simulated/measured CD maxima for the intrinsic planar chirality are 0.98/0.88, 0.99/0.93, 0.99/0.92, and 0.99/0.91 with the simulated/measured Q-factors of 2457/390, 480/121, 198/62, and 128/38 for $\delta$ =20, 40, 60, and 80 nm, respectively. Such simultaneously achieved strong chirality and high Q-factors in experiment promise chiroptical functionalities with strong light-matter interaction.

**Conclusion**

We have proposed and realized experimentally planar chiral metasurfaces with in-plane symmetry-reduced inclusions. Such metasurfaces are governed by the BIC physics with birefringence degrees of freedom and it is identified as a vortex polarization singularity surrounded by ellipse eigenstate polarizations with non-vanishing helicity. Both extrinsic and intrinsic planar chirality can be realized by symmetry breaking. By illumination symmetry breaking, the BIC state can be transferred to q-BIC with strong extrinsic chirality accompanied by tunable linewidth and signs. By introducing in-plane structure asymmetry, the intrinsic planar chirality can be achieved under normal incidence with maximum CD of 0.99 (theory) and 0.93 (experiment) and Q-factors exceeding 390 (experiment) at optical frequencies. Beyond the chirality based on circular polarizations, we could achieve any elliptical chirality by varying simultaneously the degree of birefringence and the axis angle of the anisotropic inclusion, which suggests a way to spark new research works. Due to the high Q-factor of the planar chiral q-BIC with its accessibility and controllability, our results provide unique opportunities for many applications requiring ultrahigh Q-factors and chirality control, including circular polarization emitter, arbitrary wavefront lasing, and chiral nonlinear optics.

**Methods**

*Simulation of reflection and transmission efficiency*: Rigorous coupled wave analysis ($S^4$)[59] simulations were carried out to simulate the planar chiral q-BIC metasurface and to calculate efficiencies for the co-polarization handedness-preserved coefficients, and the circular cross-polarization conversion coefficients transmission spectra. The metasurfaces are designed by α-Si columns based on a $SiO_2$ substrate.

*Fabrication of samples:* The metasurfaces were fabricated on a fused quartz substrate by utilizing electron beam lithography and including the processes such as deposition, patterning, lift-off and etching. First, a 350-nm-thick amorphous silicon (α-Si) film was deposited by plasma-enhanced chemical vapor deposition (PECVD) method at 120 °C, during which the flow rates of $SiH_4$ and Ar are 10 and 475 sccm. And the deposition pressure and RF power are 650 mTorr and 20 W, respectively. Then a PMMA film of 300 nm was spinning coated and covered by PEDOT: PSS film as a conducting layer. The desired structure was patterned by utilizing JEOL 6300FS EBL at a base dose of 1000 μC/cm² with an accelerating voltage of 100 kV for 1 hour. After the exposure process, the conducting layer was washed away and the resist was developed in 1:3 MIBK: IPA solution for 40 s and rinsed in IPA for 30 s successively, followed by a deposition of 80 nm Cr using electron beam evaporation deposition (EBD) method. To realize the lift-off process, the sample was immersed in hot acetone of 75 °C and cleaned by ultrasonic. Finally, by using inductively coupled plasma (ICP) reactive ion etching (RIE) method with HBr at room temperature (RT) for 260 s (flow rate of 50 sccm, the pressure of 10 mTorr, RF and ICP power of 50 and 750 W, respectively), the desired structure was transferred from Cr to silicon and the residual Cr was removed by cerium (IV) ammonium nitrate. The five planar chiral q-BIC metasurfaces are composed of 500×500 periods (425 μm×425 μm) with different cross-sections ($δ$ = 0 nm, 20 nm, 40 nm, 60 nm, and 80 nm), and the final structures are shown in Fig. 4a.

*Characterization of samples:* A supercontinuum laser (Fianium-WL-SC480) was employed as the broadband light source for the measurement of the transmission spectra of all polarization components under circular polarization incidences. Incident light with a circular polarization was generated by cascading a broadband polarizer and a quarter waveplate from the supercontinuum laser. Then, the incident light was focused on the sample by a lens with a focal length of 5 cm. Subsequently, the polarization components are detected by a quarter waveplate and a polarizer. The transmittances were measured by means of an Ocean spectrometer (flame-NIR). Moreover, the sample was mounted on a rotation stage for varying incident angles in the measurement.

**Author Contributions**
Z.-L.D. and T.S. conceived the idea. T.S., Z.-L.D. and X.L. designed the experiments. T.S., Z.-L.D. carried out the design and simulation of the metasurface. T.S., Z.-L.D., G. H. A.O and Y. Z conducted the theoretical analysis of the results. G.G. and J.L. fabricated the samples. T.S.

performed the measurements. Z.-L.D., Y.K., and X.L. supervised the project. Z.-L.D., T.S., and X.L. analyzed the data and wrote the manuscript. All authors contributed to the revision and discussion of the manuscript.

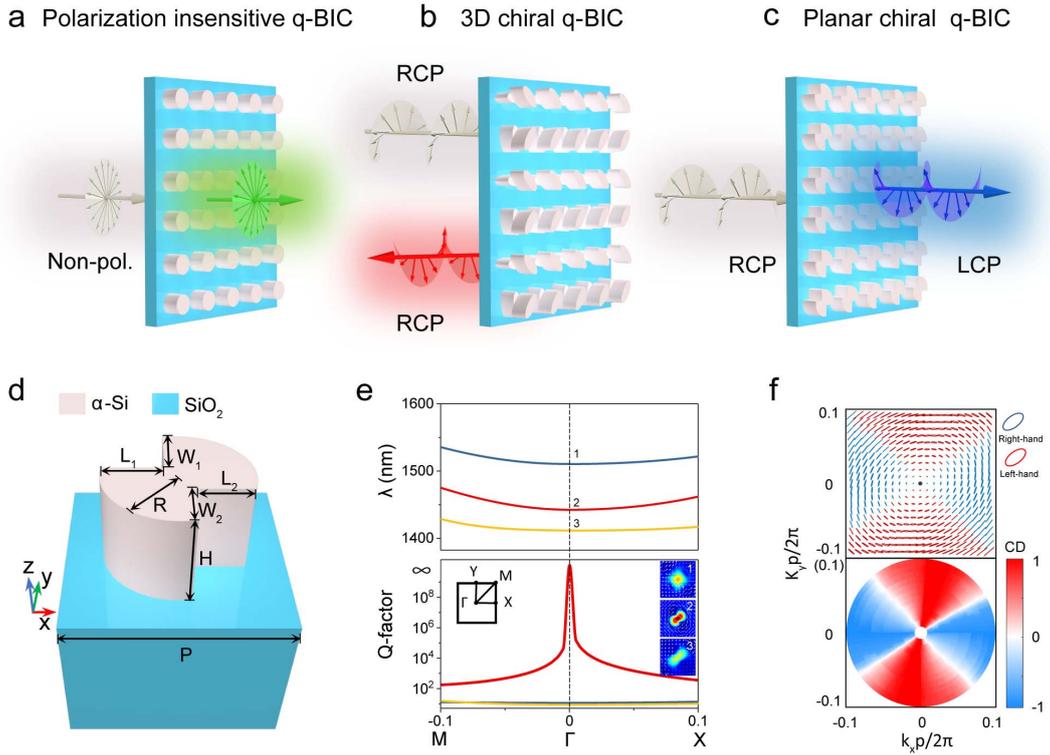

**Figure 1. Schematics of the concepts of different BIC states and the planar chiral q-BIC metasurface with a polarization singularity enclosed by vortex elliptical eigen-polarizations with non-vanishing helicities.** (a-c) Comparison of q-BICs (a) in planar metasurfaces without polarization effect, (b) in stereo-nanostructures with 3D chiral effect in reflection, and (c) in planar metasurfaces with 2D chiral effect in transmission. (d) Schematic of the unit cell of the planar chiral q-BIC metasurface formed by the amorphous silicon (α-Si) DSS structure on a SiO$_2$ substrate with parameters $P$=850 nm, $R$=280 nm, $L_1$=$L_2$=220 nm, $W_1$=$W_2$=191 nm, and $H$=350 nm. (e) Band structure (upper panel: real part of the eigen-wavelength, lower panel: Q-factors of the eigenmode) of the DSS metasurface near the Γ point. Insets show the first Brillouin zone and field patterns (color: magnetic distributions, arrow: electric vectors) of the tree typical modes. (f) Eigen-polarization profiles and the CD map in the *k*-space in the vicinity of the Γ point. On the eigen-polarization profiles, the polarization states are represented by ellipses of which the red and blue represent the eigen left-handed states and right-handed states, respectively, and the black dot represents the *V* point (BIC).

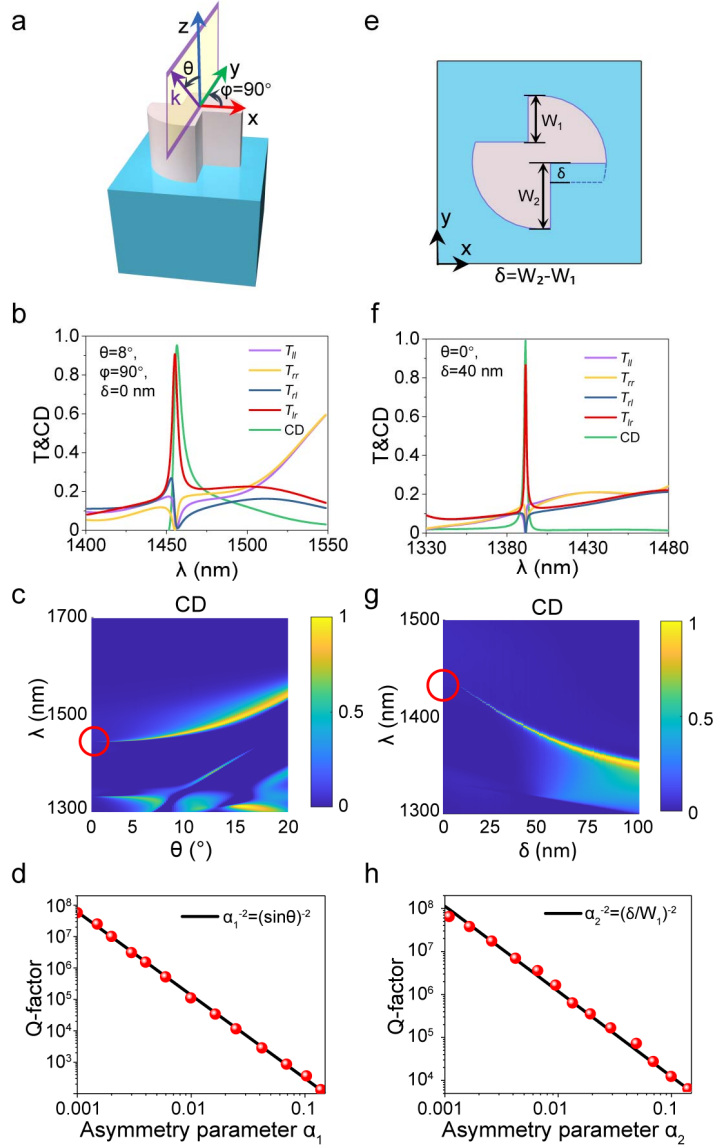

**Figure 2. Proposed implementation of the DSS structure with both the extrinsic and intrinsic planar chirality by transferring the BIC to q-BICs.** (a, e) Schematic of the symmetry-breaking processes that transfer the BIC to planar chiral q-BICs: (a) illumination symmetry breaking by varying incident angle $\theta$, (e) in-plane geometry symmetry breaking by $\delta=W_2-W_1$. (b, f) Simulated transmission Jones matrix spectra of $T_{ll}$, $T_{rr}$, $T_{rl}$ and $T_{lr}$ as well as the CD spectrum of the metasurface with (b) the same parameters as Fig. 1d ($\delta = 0$ nm) under oblique incidence ($\theta=8°$, $\varphi=90°$) and with (f) an asymmetric structure parameter ($\delta =40$ nm) for normal incidences. (c, g) The evolution of CD spectra by continuous varying (c) incident angle $\theta$ along the $\varphi=90°$ direction and by (g) the geometry asymmetry parameter $\delta$. (d, h) Dependence of the Q-factors of the planar chiral q-BIC mode on the relative asymmetry parameter (d) $\alpha_1=\sin\theta$, (h) $\alpha_2=\delta/W_1$ around the chiral q-BIC state. The solid line shows an inverse quadratic fitting.

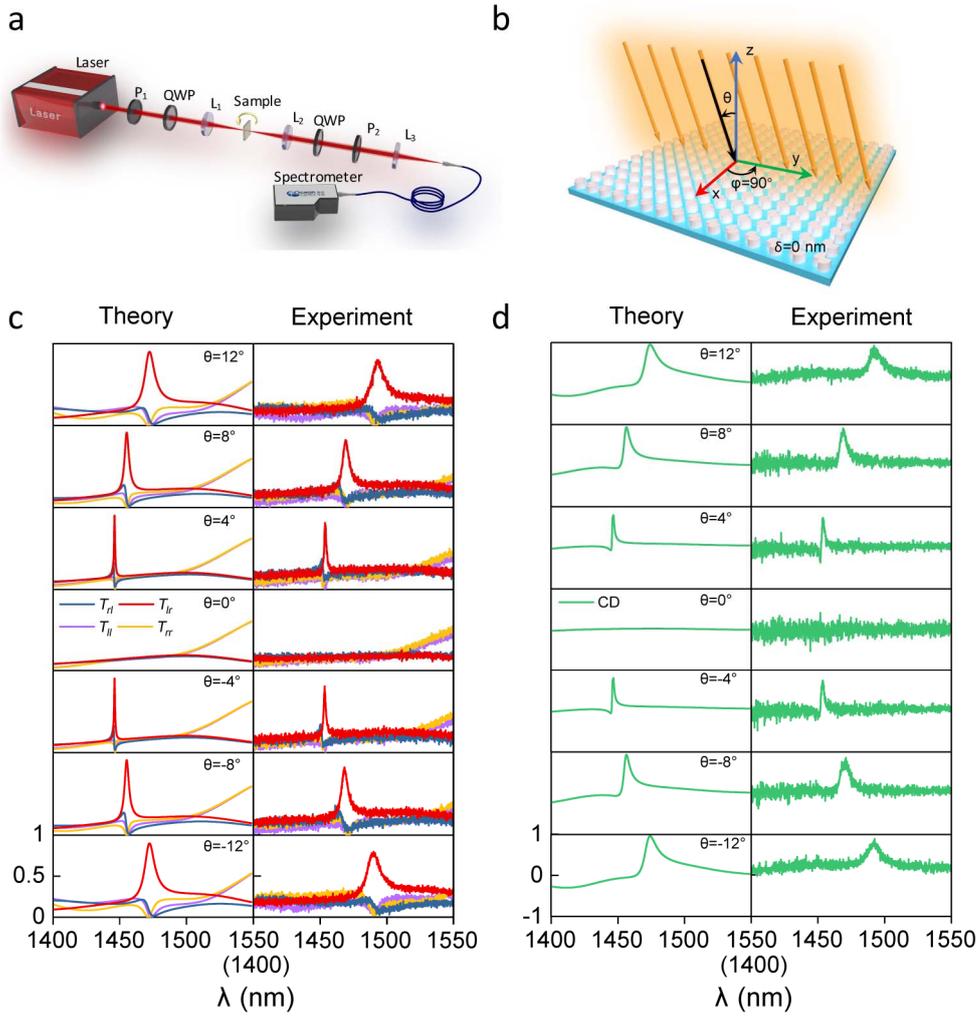

**Figure 3. Experimental verification of the extrinsic planar chiral q-BIC with illumination symmetry breaking.** (a) Experimental setup for the Jones matrix spectra measurement under the circular polarization basis. $P_1$ and $P_2$ represent the polarizers, QWP is the quarter-wave plate, $L_1$, $L_2$, and $L_3$ are lenses. (b) Schematic of the metasurface with $\delta=0$ supporting chiral q-BICs through different incident angles along the $\varphi=90°$ direction. (c) Simulated and measured transmission Jones matrix spectra ($T_{lr}$, $T_{rl}$, $T_{rr}$ and $T_{ll}$) for different incident angles ($\theta=0°$, $\pm4°$, $\pm8°$, $\pm12°$) along the $\varphi=90°$ direction. (d) Simulated and measured CD spectra for different incident angles extracted from the Jones matrix spectra in (c).

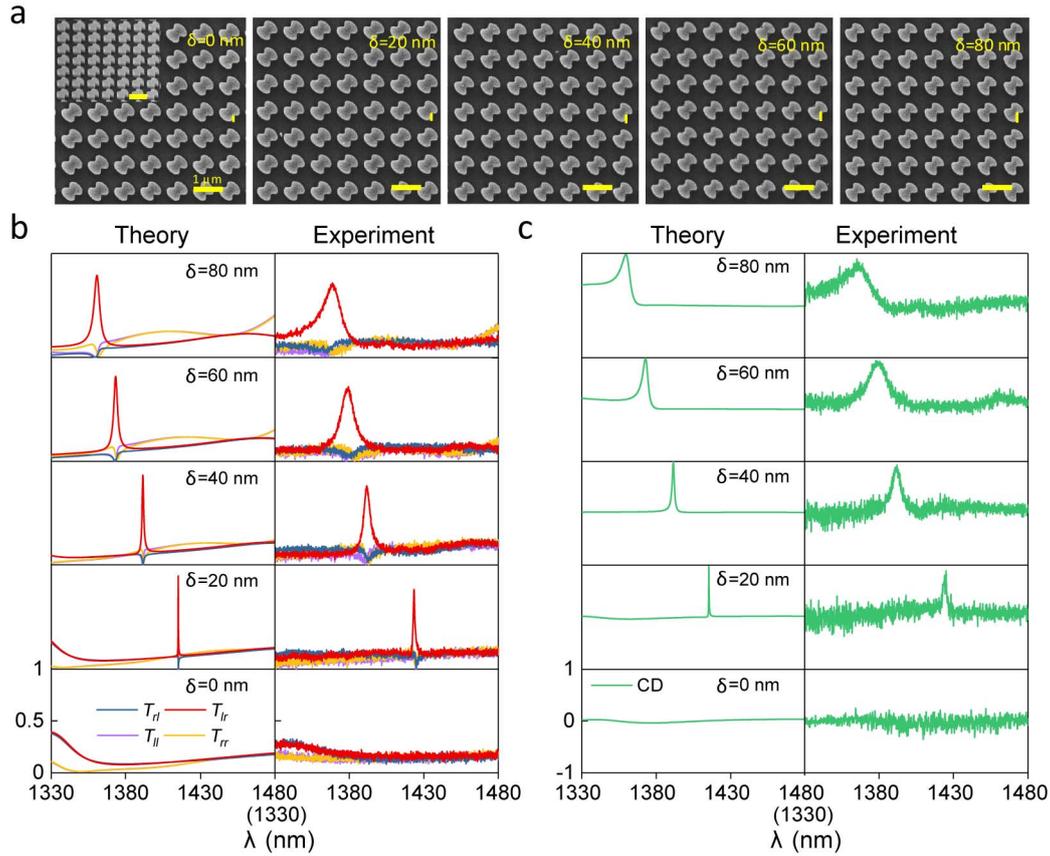

**Figure 4. Experimental verification of the intrinsic planar chiral q-BIC with various in-plane geometry asymmetry parameters.** (a) The top view of the SEM image of five fabricated DSS metasurfaces with asymmetry parameters $\delta$=0 nm, 20 nm, 40 nm, 60 nm, and 80 nm, respectively. The inset shows the slant view of the metasurface. Scale bar: 1 μm. (b) Simulated and measured transmission Jones matrix spectra ($T_{lr}$, $T_{rl}$, $T_{rr}$ and $T_{ll}$) of the five DSS metasurface samples with different $\delta$. (c) Simulated and measured CD spectra for different $\delta$ extracted from the Jones matrix spectra in (b).